\documentstyle[preprint,revtex]{aps}
\begin{document}

\def\hr{{\hat\rho}}
\def\L{{\cal L}}
\def\efy{effectively }
\def\ef{effective }
\def\r{\rangle}
\def\l{\langle}
\def\be{\begin{equation}}
\def\ee{\end{equation}}
\def\x{\xi}

\begin{title}
INTRINSIC MECHANISM FOR ENTROPY CHANGE IN \\
CLASSICAL AND QUANTUM EVOLUTION
\end{title}

\author{E. Eisenberg$^1$ and L.P. Horwitz$^{1,2}$}

\begin{instit}
$^1$ Department of Physics, Bar-Ilan University \\
Ramat-Gan 52900, Israel. \\ \\
$^2$ School of Physics, Raymond and Beverly Sackler Faculty of
Exact Sciences\\
Tel-Aviv University, Ramat-Aviv, Israel.
\end{instit}

\begin{abstract}
It is shown that the existence of a time
 operator in the Liouville space
representation of both classical and
 quantum evolution provides a mechanism
for \ef entropy change of physical states. In particular,
an initially effectively pure state can evolve under
the usual unitary evolution to an effectively mixed state.
\end{abstract}


\narrowtext
\newpage

\section{Introduction}
\smallskip
\par The Hamiltonian evolution of states in classical
mechanics is known by the Liouville theorem to be
non-mixing, i.e., to preserve the entropy
 of the system\cite{yvon}.
 The same property holds for the quantum evolution
as well, and follows from the unitarity of the
evolution operator. Thus, in both classical and quantum
mechanics, the entropy of a (closed) system is rigorously
a constant of motion. This has been an obstacle to the
consistent description of irreversible processes from
first principles \cite{ap}. The usual use of techniques of
coarse-graining or truncation to achieve a realization
of the second law does not follow from basic dynamical
laws, and is fundamentally not consistent with the
underlying Hamiltonian dynamical structure \cite{nesm}

It is often argued that large systems appear to exhibit
irreversible behavior simply because it is impossible
to observe the precise state of
 the system \cite{wz79,zur,scw89,say}.
Thus, a partial
trace over the not observed degrees of freedom is performed,
which leads to an effective coarse-graining. However, the
exponential decay of a small unstable system, such as an atom
in an excited state, or an unstable particle (e.g. a neutron),
is also not consistent with the prediction of
the (reversible) Wigner-Weisskopf description of decay systems,
which is non-exponential at short (and long) times \cite
{khal,mis-sud,lhr}.
This short time behavior can be shown to be due to the so-called
regeneration terms, which would not appear if the description
were truly irreversible \cite{mis-sin,exner,alh}.
It thus appears that the
intrinsic reversibility implemented by the unitarity of the
evolution operator is fundamentally not consistent with
the irreversiblity which is the basis not only for statistical
mechanics, but also for the description of microscopic unstable
systems, and measurement theory.
Many attempts have been made in recent years to solve this
fundamental problem \cite{nesm,rim,ghir,np93,top}.

In this work we study one of the aspects of this problem;
we show that the existence of a ``time''
operator $T$ in the framework of Liouville space theory
\cite {misra} provides a natural and consistent
mechanism for which pure states become mixed during the evolution
 (this operator is not the actual time, but is a function
on the Liouville measure space which
 translates linearly with time).
The Liouville space is essential for this construction,
since a time operator does not exist in the usual Hilbert
space (for a semi-bounded Hamiltonian).

The notion of a pure state is defined by means of
expectation values of observables, i.e., a state is
called ``pure'' if the expectation value of each
observable in this state is equal to the corresponding
expectation value computed with respect to some
well-defined wave function (defined up to a phase),
 that is, a density matrix which
is a projection operator to a one-dimensional subspace.
 In the following, we consider
the subset of a complete set of observables defined on the
Liouville space which are not explicitly dependent on the
$T$-variable. It is shown that this subset corresponds
to the experimentally accessible observables.
One obtains all the physical
information concerning this subset of observables from
an effective state resulting from the reduction of the
full state by integration over the degree of freedom
which is not relevant for this subset, i.e., the
spectrum of the {\it time operator}. We call this reduced
state the {\it \ef} physical state. It is, of course,
an old technique of statistical mechanics to trace over
unmeasured variables to obtain a reduced density matrix.
We show that the selection of the time variable in the
Liouville space is natural and appropriate, and provides
a natural mechanism for which an initially pure \ef
state can evolve to a mixed one, under the time
evolution of the system.
We show that there
exist mixed states for which the \ef physical state is
pure and denote them as \efy pure. These states may
become \efy mixed during the evolution of the system.
We formulate these ideas in the framework of the
quantum Liouville space, and consider later their
application to classical mechanics. We also consider a
simple explicit example to illustrate this mechanism.
\section{ Representation in terms of Liouville time}
\par It is well known that one cannot define a time-operator
$T$ in the usual quantum Hilbert space with semi-bounded
Hamiltonian \cite{gqm}; however, it is possible to define such
an operator in the framework of Liouville space in
which the generator of the evolution is the
``Liouvillian'' (whose prototype is the commutator with
the Hamiltonian) which generally has absolutely
continuous spectrum on the whole real axis \cite{misra}. It
is defined by
\be
\label{defL}
	 e^{-i\L t}\rho\,=\,e^{-iH t}\rho
				  e^{iH t}\,,\ee and
	 $\L_0$ by \be e^{-i\L_0 t}\rho\,=\,
				      e^{-iH_0 t}\rho
	 e^{iH_0 t}\,.
\ee
Then, $\L_I\equiv\L-\L_0$.
\par The existence of the time-operator has been extensively
used in the context of quantum statistical mechanics
\cite{top}. The kernel representing a Hilbert-Schmidt operator
$A$ on the original Hilbert space of $n$ degrees of
freedom, $\l{\bf k}|A|{\bf k'}\r$, where ${\bf k}$
consists of $n$ parameters, corresponds to the function
$A({\bf k},{\bf k'})\equiv\l{\bf k},{\bf k'}|A\r$
representing the vector $A$ of the Liouville space. We
then change variables from ${\bf k},{\bf k'}$ to $\x$,
the spectrum of $T$, and $(2n-1)$ other independent
parameters $\beta$. This transformation is defined by a
kernel $K(\x,\beta | {\bf k},{\bf k}')$ such that
\be
A(\x,\beta)\,\equiv\,\l\x,\beta|A\r\,=\,\int K(\x,\beta
| {\bf k},{\bf k}') \l {\bf k},{\bf k}'|A\r d{\bf
k}d{\bf k}'\,.
\ee
In what follows, we shall use the time operator $T$ conjugate
to the unperturbed Liouville operator $\L_0$, which is defined
according to the decomposition
\be
\label{decom}
\L\,=\,\L_0\,+\,\L_I\,,
\ee
i.e., on a suitable domain
\be
\label{commut}
[T,\L_0]=i\,.
\ee
It follows from (\ref{commut}) that
\be
\label{freeev}
\bigl(e^{-i\L_0 t}A\bigr)(\x,\beta)\,=\,A(\x+t,\beta)\,.
\ee
{}From (\ref{defL}) and (\ref{freeev}), we see that if the
free Hamiltonian is diagonal in the (generalized) states
$\{\vert {\bf k} \rangle\}$,
\be
 K(\xi + t, \beta \vert {\bf k,k'}) =
K( \xi, \beta \vert {\bf k,k'}) e^{-i(E_{\bf k} -E_{\bf k'})t},
\ee
where $E_{\bf k}$ is the unperturbed energy associated with
the variables ${\bf k}$, and hence that
\be
\label{redK}
  K(\xi , \beta \vert {\bf k,k'}) = K(0, \beta \vert {\bf k,k'})
e^{-i(E_{\bf k} - E_{\bf k'}) \xi}.
\ee
Under the free evolution, the representation of $A$ on
the Liouville space undergoes translation $\x\to\x+t$,
so that translation in the spectrum of the $T$-operator of
the Liouville space reflects the  interval of the free
evolution. We emphasize that $\x$ is not the {\it time},
but a function on the manifold of the Liouville space
(which does not exist on the manifold of the Hilbert
space), and shifts translationally with the time $t$ under
free evolution.

Using this new basis, the expectation value of an
observable is written as
\be
\label{ev}
\l A\r_\rho\,=\,{\rm Tr}(A\rho)\,
=\,\int\rho_\x(\beta)A(\x,\beta)d\x d\beta\,,
\ee
where
\be
\label{trans}
A(\x,\beta)\,=\,\int K(\x,\beta |{\bf k},{\bf
k}') \l {\bf k},{\bf k}'|A\r d{\bf k}d{\bf k}'\,.
\ee

It follows from Eq. (\ref{freeev}) that $\x$-independent
observables commute with the free Hamiltonian $H_0$. In
this case, clearly the asymptotic form of the
observable $A$ (in Heisenberg picture) exists if the
wave operator for the scattering theory exists, i.e.,
\begin{eqnarray}
\lim_{t\to\pm\infty}e^{-i\L t}A &=&
\lim_{t\to\pm\infty} U(t)^{-1}AU(t)\nonumber \\
&=&\lim_{t\to\pm\infty}U(t)^{-1}U_0(t)AU_0(t)^{-1}U(t)\nonumber \\
&=&\Omega_\pm A \Omega_\pm^{-1} = A_\pm,
\end{eqnarray}
where $U(t)$ is the full
evolution operator, and $U_0(t)$ is that of the
unperturbed evolution \cite{rem}. The $\x$-independent
observables therefore have a correspondence with the
asymptotic variables in a scattering theory \cite{amm}.
Ludwig \cite{lud} has emphasized that measurements on a
quantum system are made by means of the detection
of signals corresponding to observables which are
operationally on a semi-classical or classical level.
These measurable signals, which characterize the state
are the properties propagating to the
detectors, and are therefore asymptotic variables,
i.e., $\x$-independent. We do not argue that observables
which are time dependent in Heisenberg picture (such as
the electro-magnetic field) play no role. These operators
may be even useful for calculations of measurable quantities,
and their expectation values can be evaluated
using, for example, the Schwinger-Keldysh technique \cite{keld}.
However, from a physical point
of view, based on the above mentioned theoretical arguments
on the nature of measurement, only
functions of these observables  which have asymptotic limits
(in the case of electro-magnetic field, the free number
density and the momentum, for example) provide for
experimental measurement.  Measurements carried out upon
an evolving system involve, in fact, interactions with
apparatus which are essentially asymptotic (e.g.
magnetic fields far from an electron beam, or the $e$-$\nu$
or photon signal from the pions in the final state of K-meson
decay). These asymptotic observables determine the structure
of the state, and hence (with a sufficient number of such
measurements) can be used to define the
nature of the evolution, i.e., whether a pure state tends to a
mixed state. We thus conclude that the subset of $\x$-independent
observables corresponds to all the experimentally accessible
measurements, and is therefore the subset of observables which
can be used to characterize experimentally the structure of
a physical state.

\section{Effective states}

In view of the ideas presented in the previous section, one sees
that since the $\x$-dependent observables are experimentally
unmeasurable, they form natural candidates for the reduction
of the density matrix through integrating over the unmeasurable
variables. Note, however, that in this version of the reduction scheme
the set of unmeasured variables is not chosen arbitrarily
by specifying the macroscopic measured quantities, but it is rather
obtained through a fundamental theoretical argument concerning the
nature of the measurement process. Thus, this reduction does not
depend on the choice  of the set of variables used for the
characterization of the physical state in some specific experiment,
but is rather determined by the set of all measurable variables.
Moreover, this reduction applies to unstable microscopic
systems as well.

One thus looks for a partial
trace over $\rho$, such that the information concerning expectation
values of all the $\x$-independent observables can be extracted
from the reduced density matrix $\hr$. We now show that this desired
reduction is obtained by integrating $\rho$ over $\x$.
If $A$ belongs to the subset of
$\x$-independent operators, i.e., $A(\x,\beta)\equiv
A(\beta)$, then from Eq. (\ref{ev}) it follows that
\be
\l A\r\,=\,{\rm Tr}(A\rho)\,=\,\int\hr(\beta)A(\beta)d\beta\,,
\ee
where $\hr$ is defined as
\be
\label{defhr}
\hr(\beta)\equiv\int d\x
\rho_\x(\beta)\,.
\ee
It is therefore clear that with
respect to the set of $\x$-independent observables, all
of the information available in the state is contained
in $\hr$.

We call a state $\hr$  {\it\efy pure} if there exists
a wave function $\psi$ such that for every
$\x$-independent observable $A$
\be
\label{pcon}
\l\psi|A|\psi\r\,=\,\l
A\r_{\hr}\,=\,\int
\hr(\beta)A(\beta)d\beta\,.
\ee
According to the preceding discussion, the expectation value
of experimentally measurable observables can be predicted in
this case by means of a wave-function.
Consequently, an \efy pure state can not be experimentally
distinguished from a state which is described by a wave-function
in the usual quantum-mechanical Hilbert space (a pure state). However,
as we shall see, this \ef purity may not be maintained in time.

If $\rho$ is pure in the usual sense, i.e., ${\rm Tr}
\rho^2 = 1$, then the condition (\ref{pcon}) holds for any
observable, and therefore the resulting $\hr$ is \efy
pure. On the other hand, it is clear that the reduction
of Eq. (\ref{defhr}) is not one to one and therefore each $\hr$
corresponds to an {\it equivalence class} of states in
Liouville space. Even if only one of these states is
pure, $\hr$ would be \efy pure, since it does not
distinguish between elements of the equivalence class.
A more precise characterization of the \efy pure
states can be found in Appendix A.
It is easy to demonstrate that
$\hr(\beta)$ is uniquely determined by the measurement of
all the $T$-independent observables.
We thus see that strict purity implies \ef purity but
not the opposite, i.e., {\it even mixed states may
appear as \efy pure}.

As an illustration, consider an unstable atomic state, which
decays due to coupling with the electro-magnetic field.
The unstable state itself, is a pure state, described by a
well defined wave-function in the quantum-mechanical Hilbert
space (which is an eigenfunction of the unperturbed atomic
Hamiltonian). As discussed in the introduction, the unitarity of
the quantum evolution implies that the purity of the state is
preserved, and its entropy is not changed. However, as we have
just shown, there exist many {\it mixed} states which are
{\it identical} to the unstable state from experimental point of
view, since no accessible measurement (i.e., measurement of an
observable which commutes with the atomic free Hamiltonian, such as
the unperturbed energy or the angular momentum)
can distinguish between
these mixed state and the unstable pure state. These states
were just defined as \efy pure. As we shall see, the coupling
to the external field may, in general, induce mixing, i.e.,
the \efy pure states may evolve into \efy mixed states. This
mechanism is demonstrated for a simple example later, and
will be discussed elsewhere in the context of the unstable atomic
state.

\section{Dynamical evolution}

We wish to show now that while unitarity excludes the
possibility of mixing of pure states, mixing of \efy
pure states (destruction of the effectively pure property)
is still possible. Generally, in the
presence of interaction, the full Liouvillian takes the
form (from (\ref{decom}) and (\ref{commut}))
\be
\l\x|\L|\x'\r\,=\,-i\partial_\x\delta(\x-\x')\,+\,\l
\x|\L_I|\x'\r\,,
\ee
where the second term is, in general, not
diagonal, but rather acts as an integral operator on
$\x$. Such an evolution generator was discussed recently in
connection with the quantum Lax-Phillips theory \cite{pp}.
The resulting evolution is also of an integral operator
structure and takes the form
\be
\label{genev}
\rho_\x^{t}\,=\,\int W_{\x,\x'}(t)\rho_{\x'}^0 d\x'\,,
\ee
where the
operator $W_{\x,\x'}(t)$ acts only on the $\beta$
dependence.

For simplicity we use the Fourier transform
representation
\be
\rho(\alpha,\beta)\,=\,\int
e^{-i\x\alpha}\rho_\x(\beta) d\x ;
\ee
\be
{\overline
W}_{\alpha,\alpha'}(t)\,=\,\int e^{-i\x\alpha}
e^{i\x'\alpha'}W_{\x,\x'}(t)d\x d\x'\,.
\ee
Note that $\hr(\beta)=\rho(\alpha,\beta)|_{\alpha=0}$,
and therefore
\be
\label{evol}
\hr^t(\beta)=\rho^t(\alpha,\beta)|
_{\alpha=0}=\int{\overline
W}_{0,\alpha'}(t;\beta,\beta')\rho(\alpha',\beta')
d\alpha'd\beta'\,.
\ee
The initial \ef purity of $\rho$ provides information
only on its $\alpha=0$ component while the other
components may be even \efy mixed, but, as we see from
Eq. (\ref{evol}), during the evolution the $\alpha=0$ component
develops contributions from the other components, and
therefore it may become mixed. The states keep their
effective purity, in general, only if ${\overline
W}_{0,\alpha'}\sim\delta(\alpha')$.

We wish to consider now a simple concrete example to
illustrate the above ideas. Consider the evolution of
a particle in three dimensions in the presence of a
screened Coulomb (Yukawa) potential. The matrix
elements of the free Liouvillian are given by (we take
$2m=1$)
\be
\l {\bf k}_1,{\bf k}_2|\L_0|{\bf k}_3,{\bf
k}_4\r=\delta^3({\bf k}_1-{\bf k}_3)\delta^3({\bf k}_2-{\bf
k}_4)({\bf k}_2^2-{\bf k}_1^2)\,.
\ee
We change the variables in Liouville space from $({\bf
k}_1,{\bf k}_2)$ to $(\alpha,
{\bar \beta},\Omega_1,\Omega_2)$ by
the transformation
\be
\alpha={\bf k}_2^2-{\bf k}_1^2,\qquad {\bar \beta}={\bf
k}_2^2+{\bf k}_1^2,
\ee
and $\Omega_1,\Omega_2$ are the angle variables of the
momenta ${\bf k}_1,{\bf k}_2$, respectively. We denote
the set of variables ${\bar \beta},\Omega_1,\Omega_2$ by
$\beta$. In this new basis the matrix elements of
the free Liouvillian are given by
\be
\l \alpha,\beta|\L_0|\alpha',\beta'\r=
\alpha\delta(\alpha-\alpha')\delta(\beta-\beta')\,.
\ee
The variables $\alpha, {\beta}$
 defined by this change of basis coincide with
the $\alpha$,$\beta$ of our general discussion above.

As mentioned before, \efy pure states are mixed during
the evolution unless ${\overline
W}_{0,\alpha'}\sim\delta(\alpha')$. We therefore look
at the evolution operators induced by the perturbation
to see whether this is the case. The matrix  elements
of the interaction Liouvillian are given by
\be
\l {\bf k}_1,{\bf k}_2|\L_I|{\bf k}_3,{\bf k}_4\r=\delta^3
({\bf k}_1-{\bf k}_3){\tilde V}_{{\bf k}_2-{\bf k}_4}
-\delta^3({\bf k}_2-{\bf k}_4){\tilde V}_{{\bf k}_1-{\bf k}_3}\,,
\ee
where ${\tilde V}_{\bf
k}$ is the Fourier transform of the potential $V$,
taken at the point ${\bf k}$.

For the screened Coulomb potential
\be
V(r)={Ae^{-\mu r}\over \mu r}\,,
\ee
${\tilde V}_{\bf k}$ is given by
\be
{\tilde V}_{\bf k}={4\pi A\over \mu({\bf
k}^2+\mu^2)}\,,
\ee
and the matrix element takes the form
\be
\l {\bf k}_1,{\bf k}_2|\L_I|{\bf k}_3,{\bf k}_4\r={4\pi A\over\mu}
\Biggl({\delta^3({\bf k}_1-{\bf k}_3)\over({\bf k}_2-{\bf
k}_4)^2+\mu^2}-{\delta^3({\bf k}_2-{\bf k}_4)\over({\bf
k}_1-{\bf k}_3)^2+\mu^2}\Biggr)\,.
\ee
Changing the variables to $(\alpha,\beta)$, one obtains
\begin{eqnarray}
\label{examp}
\lefteqn{\l\alpha,\beta|\L_I|\alpha',\beta'\r\equiv
\L_I(\alpha,\alpha',\beta,\beta')=}\nonumber \\
& &{64\pi A\over\sqrt2\mu}\Biggl({\Bigl[
\delta({\bar \beta}-\alpha-
{\bar \beta}'+\alpha'){1\over\sqrt{{\bar \beta}
-\alpha}}\Bigr]\delta(\Omega_1,\Omega_3)\over
{\bar\beta}+{\bar\beta}'+\alpha+\alpha'-2\sqrt{({\bar\beta}+\alpha)
({\bar\beta}'+\alpha')}B(\Omega_2,\Omega_4)+\mu^2}\nonumber \\
& & -{\Bigl[\delta({\bar\beta}+\alpha-{\bar\beta}'-\alpha')
{1\over\sqrt{{\bar\beta}+\alpha}}\Bigr]\delta(\Omega_2,\Omega_4)
\over({\bar\beta}+{\bar\beta}'-(\alpha+\alpha')-
2\sqrt{({\bar\beta}-\alpha)
({\bar\beta}'-\alpha')}B(\Omega_1,\Omega_3)+\mu^2}\Biggr),
\end{eqnarray}
where $B(\Omega_1,\Omega_2)$ is defined by
\be
B(\Omega_1,\Omega_2)=\sin\theta_1\sin\theta_2
\cos(\phi_1-\phi_2)
+\cos\theta_1\cos\theta_2\,.
\ee

It is therefore clear that the kernel \
$\L_I(\alpha,\alpha',\beta,\beta')$ \
is {\it not} of the form
$\delta(\alpha-\alpha')\hat A
(\beta,\beta')$ and therefore
the evolution operators do not have this form either.
In particular, for weak interactions, first order
perturbation theory gives
\be
\label{ex:mix}
{\overline
W}_{0,\alpha}(t;\beta,\beta')=\delta(\alpha)
\delta(\beta - \beta')
-it\L_I(0,\alpha,\beta,\beta')+O(t^2A^2)\,,
\ee
where the second term induces mixing.

We have shown that no mixing occurs if the unperturbed
Liouvillian is non-degenerate (see Appendix A), and the result
(\ref{ex:mix}) shows that no mixing occurs for the free motion.
We do not yet have a general classification.

\section{Entropy}

We next define the notion of {\it entropy} for the \ef
states and show that this entropy is not constant during the
motion as in traditional quantum (and classical) mechanics.

We now remark that since, according to Eq.
(\ref{redK}), the density operator
in $(\x,\beta)$ representation can be written as
\be
\rho_\x(\beta)=\int K(0,\beta \vert
{\bf k},{\bf k'})e^{-i(E_k-E_{k'})\x}
\rho({\bf k},{\bf k'}),
\ee
one obtains
\be
\hr(\beta)=\int d\x \rho_\x(\beta) = 2\pi\int K(0,\beta \vert
{\bf k},{\bf k'})
\delta(E_k-E_{k'})\rho({\bf k},{\bf k'}).
\ee
The necessary and sufficient condition
for a state to be \efy pure, as pointed
out in Appendix A, is that the function $\rho({\bf k},{\bf k'})$
be factorizable in the equal energy subspaces.

The entropy of a quantum system, defined as
\be
S=-{\rm Tr}\rho\ln\rho
\ee
to satisfy the requirements of convexity and
additivity, vanishes for a pure
state, i.e., a density operator of the form
 $\rho=\vert\psi\r\l\psi\vert$, where
the norm squared $\l\psi\vert\psi\r$ is unity.
We therefore define the entropy
of the \ef state as the sum of entropies associated with
$\rho({\bf k},{\bf k'})$ in each
energy subspace, i.e., with the reduced
operator
\be
\hr =\int d{\bf k}d{\bf k'}\delta(E_k-E_{k'})\rho({\bf k},{\bf k'})
\vert{\bf k}\r\l{\bf k'}\vert.
\ee

For simplicity, we defer to Appendix B all the precise
mathematical details required for the above reduction,
and state here only the result
\be
S = \int dE S_E = -\int dE\ {\rm Tr}\hr_E\ln\hr_E .
\ee
where $\hr_E$ is the (normalized; see App. B) density matrix restricted
to the $E$-energy subspace. In case $\rho(E,\Omega_k;E,\Omega_{k'})$
is factorizable, $S_E$ vanishes. Hence, for \efy pure $\rho$, where
this factorizability condition holds for each equal energy subspace,
the entropy is zero. However, in case $\rho$ is not \efy pure,
since $S$ is convex, the admixture of non-factorizable
elements inside the equal energy subspaces (as induced by the evolution
(\ref{ex:mix})) induces an increase of entropy.
One therefore sees that the entropy is {\it increased} in course of
the evolution for a general \efy pure initial state unless it is a
strictly pure state, i.e., described by a wave-function, whose entropy
is constant. We do not study here the
conditions under which the entropy increases
given some non-zero entropy initial state.

\section{Concluding remarks}

We have shown that the class of observables which
are constants of the free motion determine a reduced density
matrix which, even when the original density matrix of the
system corresponds to a mixed state, may be \efy pure. Such
states correspond to an equivalence class which include,
therefore, both pure and mixed states.  An equivalence class
of \efy pure states contains only pure states if and only
if it is non-zero in only one energy subspace; due to the
trace condition, such an
equivalence class can be realized on discrete spectrum.
\par  Under
the evolution of the system, an \efy pure state
 may become \efy mixed,
i.e., the elements in equal energy subspaces may become
non-factorizable.  We have defined the entropy of such a system
which vanishes for an \efy pure state. We have shown that the
system may evolve from any initial \ef state
to an \ef state for which the entropy has changed.
In particular, if the initial state is \efy pure, evolution
can lead to an \efy mixed state with non-vanishing entropy.
The example
of the screened Coulomb potential which we worked out here
illustrates this effect, and furthermore shows explicitly
that the free evolution does not change the entropy.

The method that we have described above applies as well
to the formulation of classical mechanics on a Hilbert
space defined on the manifold of phase space which was
introduced by Koopman \cite{koop}  and used extensively in
statistical mechanics \cite{nesm}. Misra \cite{misra} has shown that
dynamical systems which admits a Lyapunov operator
necessarily have absolutely continuous spectrum;
therefore one can construct a time operator on the
classical Liouville space for such systems. We identify
the variables ${\bf k},{\bf k'}$ with the variables of
the classical phase space, and consider the trace as an
integral over this space. The expectation value of a
$\xi$-independent operator defines a reduced density
function in the form (\ref{defhr}). Since a pure state is defined
by a density function concentrated at a point of the
phase space, a state which is \efy pure must have the
form $\hr(\beta)=\delta(\beta-\beta_0)$. The
equivalence class associated with this reduced density
contains mixed states as well, such as
$\rho(\x,\beta)=\delta(\beta-\beta_0) f(\x)$
corresponding to a non-localized function on the phase
space $({\bf k},{\bf k}')$. The structure of the
theory, and the conclusions we have reached, are
therefore identical to those of the quantum case.

\newpage
\appendix{}

In this appendix we intend to characterize an \efy pure
state more explicitly. Since $\alpha$ is the Fourier
dual of the variable $\x$, which is the spectrum of the
$T$-operator, it follows from (\ref{commut}) that
$\alpha$ is the spectrum of the unperturbed Liouvillian
$\L_0$, the canonical conjugate of $T$. Hence, the $\alpha=0$
component of a state
$\int d{\bf k}d{\bf k'} c({\bf k},{\bf k}')\vert{\bf k}\rangle
\langle{\bf k'}\vert $,
for a basis $\{\vert{\bf k}\rangle\}$ which are (generalized)
eigenfunctions of $H_0$ with (generalized)
eigenvalues $E_{\bf k}$,
is the partial integral over the
terms for which the unperturbed Liouvillian vanishes, i.e.,
$E_{\bf k}=E_{\bf k'}$.

For a pure state corresponding to $\psi=\int
a({\bf k})\vert{\bf k}\rangle $, $c({\bf k},{\bf k}')=a({\bf
k})a({\bf k}')^*$ is factorizable. An \efy pure state
is a state which has the same reduced density matrix $\hr$
as some pure state. Since this reduction is given by
taking the $\alpha=0$ component of $\rho(\alpha,\beta)$, it
follows that the $\alpha=0$ component of an \efy pure
state coincides with the $\alpha=0$ component of a pure
state, and therefore satisfies this factorizability
condition in the equal energy subspaces. On the other
hand, this condition in the equal energy subspaces does
not imply its general validity (for $\alpha\neq 0$).
Hence, an \efy pure state is associated with an
equivalence class which includes mixed states as well.

As an example, note that the mixed state
\be
\rho = \int dE dE' \mu(E,E')
|\phi_E\rangle\langle\phi_{E'}|,
\ee
where $\mu(E,E')$ is a positive kernel, $\int dE\mu(E,E)=1$,
and the generalized states $|\phi_E\rangle$ correspond to
normalized elements of ${\cal H}_E$ (and for which $H_0$ is
a multiplication operator), is \efy pure.

If $H_0$ is nondegenerate, the \ef purity
condition holds trivially for every state (diagonal
elements of the density matrix in $H_0$ representation
are positive definite), and, with the mechanism we propose,
the evolution {\it cannot induce mixing}. Note that for
classical systems, this condition implies that the
system is integrable.

\newpage
\appendix{}

In this appendix we describe the formal mathematical reduction of
the density matrix into the equal energy subspaces.
To extract the density operator
associated with each equal energy subspace, we
define the projection density
\be
P_E = \int d{\bf k}\delta(E-E_k)\vert{\bf k}\r\l{\bf k}\vert,
\ee
and carry out the operation
\be
P_E\hr = \int d{\bf k}d{\bf k'}\delta(E-E_k)\delta(E_k-E_{k'})
\rho({\bf k},{\bf k'})\vert{\bf k}\r\l{\bf k'}\vert.
\ee
This operator is well defined; writing $d{\bf k}=dE_kd\Omega_k$,
where $\Omega_k$ is the degeneracy
manifold associated with $E_k$, one obtains
\be
\label{projhr}
P_E\hr = \int d\Omega_k d\Omega_{k'}\rho({\bf k},{\bf k'})
\vert_{E=E_k=E_{k'}}\, \vert E\Omega_k\r\l E\Omega_{k'}\vert.
\ee
The trace of this operator on the
full Hilbert space does not exist (a
well-known problem associated with continuous spectrum, and
related to the
Van Hove singularity arising from the fact
that the equilibrium state
for the unperturbed evolution is not an element of
the Hilbert-Schmidt space). We therefore consider the foliation
\be
{\cal H}=L^2({\bf R},{\cal H}_E),
\ee
with Lebesgue measure on $\bf R$, and for
which ${\cal H}_E$ corresponds to the
degeneracy subspace at each $E$ (pointwise).
 For $f\in{\cal H}$, the norm in
 this foliation is defined as
\be
\Vert f\Vert^2=\int\Vert f_E\Vert^2_{{\cal H}_E}dE,
\ee
or in terms of the original manifold $\{{\bf k}\}$,
\be
\int\Vert f_E\Vert^2_{{\cal H}_E}dE =
\int\vert f(E,\Omega_E)\vert^2 dEd\Omega_E =
\int\vert f({\bf k})\vert^2 d{\bf k} .
\ee

The trace ${\rm Tr}_E$ restricted to  ${\cal H}_E$
then corresponds to a trace
 over the degeneracy subspace alone.
 For the operator (\ref{projhr}),
\be
{\rm Tr}_E P_E \hr =
\int \rho({\bf k},{\bf k})\vert_{E=E_k}\, d\Omega_k.
\ee
Note that
\be \int dE {\rm Tr}_E P_E \hr =
\int \rho({\bf k},{\bf k})\vert_{E=E_k}\, dEd\Omega_k = 1.
\ee
We define, however,
\be
\lambda_E = {\rm Tr}_E P_E \hr
\ee
so that
\be
\hr_E = {1\over \lambda_E}P_E \hr
\ee
is a (dimensionless, normalized) operator
in the Hilbert space ${\cal H}_E$
with representation
\be
\hr_E =\int d\Omega_k d\Omega_{k'}\rho(E,\Omega_k;E,\Omega_{k'})
\vert E\Omega_k\r\l E\Omega_{k'}\vert.
\ee
The entropy is then defined as
\be
S = \int dE S_E = -\int dE\ {\rm Tr}\hr_E\ln\hr_E .
\ee


\begin{thebibliography}{99}
\bibitem{rem} In fact, it is shown in ref. \cite{amm} that asymptotic
	  observables may exist even if the usual wave
				  operators do not.
\end{thebibliography}
\end {document}